\documentclass[review]{elsarticle}

\usepackage{siunitx, subcaption}
\DeclareSIUnit[number-unit-product = {}]\linepair{lp}
\DeclareSIUnit[number-unit-product = {}]\nucleon{u}
\DeclareSIUnit[number-unit-product = {}]\radiationLength{X_0}
\newcommand{\energy}{\(E_0 \)}
\newcommand{\energyUnit}{\mega\electronvolt\per\nucleon}
\newcommand{\thickness}{\(T \)}
\newcommand{\distance}{\(D \)}
\newcommand{\clearance}{\(C \)}
\newcommand{\lengthUnit}{\milli\meter}
\newcommand{\uncertainty}{\(\sigma_\mathrm{MLP} \)}
\newcommand{\imageResolutionUnit}{\linepair\per\milli\meter}
\newcommand{\materialBudget}{\(\varepsilon \)}
\newcommand{\materialBudgetUnit}{\percent}
\newcommand{\positionResolution}{\(\sigma_\mathrm{p} \)}
\newcommand{\positionResolutionUnit}{\micro\meter}

\usepackage[dvipsnames]{xcolor}
\definecolor{TableauBlue}{HTML}{1F77B4}
\definecolor{TableauGreen}{HTML}{2CA02C}
\definecolor{TableauRed}{HTML}{D62728}
\definecolor{TableauPurple}{HTML}{9467BD}
\definecolor{TableauGrey}{HTML}{7F7F7F}

\usepackage{tikz,tikz-3dplot}
\usetikzlibrary{plotmarks}

\usepackage{hyperref}

\expandafter\let\csname equation*\endcsname\relax
\expandafter\let\csname endequation*\endcsname\relax
\usepackage{glossaries}
\newacronym{mlp}{MLP}{most likely path}
\newacronym{rmsd}{RMSD}{root-mean-square deviation}

\biboptions{sort&compress}
\makeatletter
\providecommand{\doi}[1]{%
  \begingroup
    \let\bibinfo\@secondoftwo
    \urlstyle{rm}%
    \href{https://doi.org/#1}{%
      doi:\discretionary{}{}{}%
      \nolinkurl{#1}%
    }%
  \endgroup
}
\makeatother

\newcommand{\parencite}[1]{\cite{#1}} 
\newcommand{\textcite}[1]{\citet{#1}} 

\begin{document} 
    
\begin{frontmatter}

    \title{Single particle tracking uncertainties in ion imaging}
    \journal{Physica Medica}
    
    \author[ati]{A.~Burker}
    \author[hephy]{T.~Bergauer}
    \author[ati]{A.~Hirtl\corref{mycorrespondingauthor}}
    \cortext[mycorrespondingauthor]{Corresponding author}
    \ead{albert.hirtl@tuwien.ac.at}
    \author[hephy]{C.~Irmler}
    \author[hephy]{S.~Kaser}
    \author[meduni,medaustron]{B.~Kn\"ausl}
    \author[hephy]{F.~Pitters}
    \author[hephy]{F.~Ulrich-Pur}

    \address[ati]{Atominstitut, TU Wien, 1020 Vienna, Austria}
    \address[hephy]{Institute of High Energy Physics, Austrian Academy of Sciences, 1050 Vienna, Austria}
    \address[meduni]{Division Medical Radiation Physics, Department of Radiation Oncology, Medical University of Vienna / AKH Vienna, 1090 Vienna, Austria}
    \address[medaustron]{MedAustron Center for Ion Therapy and Research, 2700 Wiener Neustadt, Austria}
    
    \date{}

    \begin{abstract}
        An extensive comparison of the path uncertainty in single particle tracking systems for ion imaging was carried out based on Monte Carlo simulations.
        The spatial resolution as function of system parameters such as geometry, detector properties and the energy of proton and helium beams was investigated to serve as a guideline for hardware developments.

        Primary particle paths were sampled within a water volume and compared to the most likely path estimate obtained from detector measurements, yielding a depth-dependent uncertainty envelope.
        The maximum uncertainty along this curve was converted to a conservative estimate of the minimal radiographic pixel spacing for a single set of parameter values.

        Simulations with various parameter settings were analysed to obtain an overview of the reachable pixel spacing as function of system parameters.
        The results were used to determine intervals of detector material budget and position resolution that yield a pixel spacing small enough for clinical dose calculation.

        To ensure a pixel spacing below \SI{2}{\lengthUnit}, the material budget of a detector should remain below \SI{0.25}{\materialBudgetUnit} for a position resolution of \SI{200}{\positionResolutionUnit} or below \SI{0.75}{\materialBudgetUnit} for a resolution of \SI{10}{\positionResolutionUnit}.
        Using protons, a sub-millimetre pixel size could not be achieved for a phantom size of \SI{300}{\lengthUnit} or at a large clearance.
        With helium ions, a sub-millimetre pixel spacing could be achieved even for a large phantom size and clearance, provided the position resolution was less than \SI{100}{\positionResolutionUnit} and material budget was below \SI{0.75}{\materialBudgetUnit}.
    \end{abstract}

    \begin{keyword}
        ion imaging, ion radiography, path uncertainty, most likely path
    \end{keyword}

\end{frontmatter}

\section{Introduction}
\label{sec:Introduction}

Recent advances in external beam radiotherapy with charged particles have led to an increased activity in ion imaging research.
Several demonstrators \parencite{Sadrozinski2013,Scaringella2014,Taylor2015,Mattiazzo2018,Esposito2018,Pettersen2019} have been developed to produce accurate three-dimensional images of the relative stopping power distribution within a patient, a quantity that is necessary for clinical treatment plan creation and dose calculation.
Single particle tracking systems in particular are able to isolate the paths and energy depositions of individual particles, yielding a better image resolution than other set-ups in ion imaging \parencite{Krah2018}.

Multiple Coulomb scattering degrades the spatial resolution of ion imaging systems because the original path through an object cannot be completely recovered \parencite{Schneider1994}.
Thus, a model that attempts to reconstruct the original path will always retain an intrinsic amount of uncertainty, even with perfectly accurate measurements.
The established standard model in an ion imaging context is the \gls{mlp} \parencite{Williams2004}.
Monte Carlo simulations have previously been used to study the intrinsic uncertainty of the \gls{mlp} model, for example, under the effects of data cuts on kink angle and energy loss \parencite{Schulte2008}, or using the energy deposition in the calorimeter to filter nuclear events in helium imaging \parencite{Volz2019}.
\gls{mlp} uncertainty has also been investigated for different ion species \parencite{CollinsFekete2017} or when taking material inhomogeneities in the phantom into account \parencite{CollinsFekete2017b,Khellaf2019,Brooke2020}.
Similarly, extrinsic parameters such as the distance between detectors \parencite{Penfold2011}, the gap between inner detectors and phantom \parencite{Schneider2012, Sadrozinski2013} or the detector material budget (detector thickness divided by its radiation length) and position resolution \parencite{Civinini2012,Sadrozinski2013} have been investigated individually.
In addition to studies based on Monte Carlo simulations, two investigations used analytical methods to estimate the uncertainty while comparing several parameters \parencite{Bopp2014} or for different types of imaging systems \parencite{Krah2018}.

System parameters influence each other and can add significantly to the intrinsic uncertainty of a reconstructed path.
For example, uncertainties due to position resolution and scattering in the detectors are more severe for a large air gap between phantom and detector \parencite{Radonic2020}.
In the authors' opinion it is therefore worthwhile to expand the parameter space beyond the limited coverage in existing literature to guide the development of a new detector system.
Some detrimental effects due to nuclear events remain even after filtering, and these are usually not considered by analytical methods.
To the best of the authors' knowledge, no systematic investigation of the \gls{mlp} uncertainty has been carried out based on Monte Carlo simulations or experimentally.

The purpose of this study is to create an extensive comparison of single particle tracking system parameters with regard to their influence on \gls{mlp} uncertainty.
Sensor properties, system geometry, phantom size and beam attributes are taken into account in the respective Monte Carlo simulations.
A minimum radiographic pixel spacing is obtained for many parameter values and their respective combinations, and is presented in an empiric summary.
The summary can be used to delimit intervals in terms of position resolution and material budget useful for ion imaging. 
It serves as a guide for upcoming hardware developments towards an imaging system at MedAustron \parencite{UlrichPur2020}, where protons and carbon ions are available with energies up to \SI{800}{\mega\electronvolt} and \SI{400}{\energyUnit}, respectively.
Moreover, a road map to establish helium ion beams is underway.

\section{Material and methods}
\label{sec:MaterialAndMethods}

A simple representation of a single particle tracking ion imaging system was modeled in Geant4 \parencite{Agostinelli2003} (section \ref{sec:MonteCarloSimulation}).
It was used to obtain the precise movement of charged particles through a phantom and two surrounding detector stations, each consisting of two tracking planes.
To emulate detector uncertainties, hits on the trackers were first convoluted with Gaussian uncertainties and then used to calculate the position and direction on the phantom surface, referred to as boundary conditions (section \ref{sec:BoundaryConditions}).
Using these boundary conditions, the path through the phantom was reconstructed by the \gls{mlp} model and the root-mean-square deviation of model positions was obtained as a function of depth (section \ref{sec:UncertaintyEnvelope}).
Simulations and analyses were carried out multiple times while iterating through sets of system parameters to investigate the influence of each parameter and to create an overview of the achievable minimum pixel spacing within the parameter space (section \ref{sec:ParameterValues}).

\subsection{Monte Carlo simulation}
\label{sec:MonteCarloSimulation}

\begin{figure}
    \centering
    \resizebox{\linewidth}{!}{
    \tdplotsetmaincoords{80}{70}
    \begin{tikzpicture}[tdplot_main_coords]

        \draw[thick,->] (0,0,0) -- (-1,0,0)
        node[anchor=south, TableauRed]{$x$} [TableauRed];
        \draw[thick,->] (0,0,0) -- (0,1,0)
        node[anchor=south east, TableauBlue]{$z$} [TableauBlue];
        \draw[thick,->] (0,0,0) -- (0,0,1)
        node[anchor=south, TableauGreen]{$y$} [TableauGreen];

        \def\rectSize{2.5}
        \def\clearanceVV{1}
        \def\clearanceV{2}
        \def\thicknessV{3}

        \def\firstPlane{-\clearanceVV-\clearanceV}
        \def\secondPlane{\firstPlane+\clearanceVV}
        \def\thirdPlane{\firstPlane+\clearanceVV+\clearanceV+\thicknessV+\clearanceV}
        \def\fourthPlane{\firstPlane+\clearanceVV+\clearanceV+\thicknessV+\clearanceVV+
            \clearanceV}

        \foreach \y in {\firstPlane,\secondPlane,\thirdPlane,\fourthPlane}
            {
                \draw[line width=0.75,dashed,TableauRed!50]
                (-\rectSize, \y, 0) -- (\rectSize, \y, 0);
                \draw[line width=0.75,dashed,TableauGreen!50]
                (0, \y, -\rectSize) -- (0, \y, \rectSize);

                \draw[line width=1.25] (-\rectSize, \y, -\rectSize) --
                (-\rectSize, \y, \rectSize) -- (\rectSize, \y, \rectSize) --
                (\rectSize, \y, -\rectSize) -- (-\rectSize, \y, -\rectSize);

                \draw[-latex] (-\rectSize,\y-0.25,\rectSize + 0.1) --
                (-\rectSize,\y-0.01,\rectSize + 0.1);
                \draw[-latex] (-\rectSize,\y+0.25,\rectSize + 0.1) --
                (-\rectSize,\y+0.01,\rectSize + 0.1) node[above]{\materialBudget};
            }

        \def\phantomStart{\firstPlane+\clearanceVV+\clearanceV}
        \def\phantomEnd{\firstPlane+\clearanceVV+\clearanceV+\thicknessV}
        \coordinate (P1) at (-\rectSize,\phantomStart,-\rectSize);
        \coordinate (P2) at (-\rectSize,\phantomStart,\rectSize);
        \coordinate (P3) at (\rectSize,\phantomStart,\rectSize);
        \coordinate (P4) at (\rectSize,\phantomStart,-\rectSize);
        \coordinate (P5) at (-\rectSize,\phantomEnd,-\rectSize);
        \coordinate (P6) at (-\rectSize,\phantomEnd,\rectSize);
        \coordinate (P7) at (\rectSize,\phantomEnd,\rectSize);
        \coordinate (P8) at (\rectSize,\phantomEnd,-\rectSize);
        \draw[line width=1.25] (P1) -- (P2) -- (P3) -- (P4) -- (P1);
        \draw[line width=1.25] (P2) -- (P6) -- (P7) -- (P3);
        \draw[line width=1.25] (P4) -- (P8) -- (P7);
        \draw[thick, dashed] (P1) -- (P5) -- (P6);
        \draw[thick, dashed] (P5) -- (P8);
        \fill[TableauGrey,opacity=0.1] (P2) -- (P6) -- (P7) -- (P3);
        \fill[TableauGrey,opacity=0.1] (P1) -- (P2) -- (P3) -- (P4);
        \fill[TableauGrey,opacity=0.2] (P3) -- (P7) -- (P8) -- (P4);
        \fill[TableauGrey,opacity=0.2] (P5) -- (P6) -- (P7) -- (P8);
        \fill[TableauGrey,opacity=0.5] (P1) -- (P5) -- (P8) -- (P4);
        \foreach \z in {1,2,3,13,14}
            {
                \draw[dotted] (-\rectSize,\z/5,-\rectSize) --
                (-\rectSize,\z/5,\rectSize) --
                (\rectSize,\z/5,\rectSize) --
                (\rectSize,\z/5,-\rectSize) --
                (-\rectSize,\z/5,-\rectSize);
            }

        \def\beamStart{\firstPlane-4}
        \def\beamEnd{\firstPlane-2}
        \draw[line width=4pt,TableauPurple,->] (0,\beamStart,0) --
        node[above,black]{\energy} node[below,black]{p, He} (0,\beamEnd,0);
        \draw[line width=2pt,white] (0,\beamStart+0.125,0) --
        (0,\beamEnd-0.125,0);

        \def\annotationX{\rectSize+1.5}
        \def\annotationY{-\rectSize}

        \foreach \z in {\firstPlane,\secondPlane,\thirdPlane,\fourthPlane,
                \phantomStart,\phantomEnd}
            {
                \draw[dotted] (\rectSize,\z,\annotationY)
                -- (\annotationX,\z,\annotationY);
            }

        \draw[latex-latex] (\annotationX,\firstPlane,\annotationY)
        -- node[below] {\distance}
        (\annotationX,\secondPlane,\annotationY);
        \draw[latex-latex] (\annotationX,\secondPlane,\annotationY)
        -- node[below] {\clearance}
        (\annotationX,\phantomStart,\annotationY);
        \draw[latex-latex] (\annotationX,\phantomStart,\annotationY)
        -- node[below] {\thickness} (\annotationX,\phantomEnd,\annotationY);
        \draw[latex-latex] (\annotationX,\phantomEnd,\annotationY)
        -- node[below] {\clearance}
        (\annotationX,\thirdPlane,\annotationY);
        \draw[latex-latex] (\annotationX,\thirdPlane,\annotationY)
        -- node[below] {\distance}
        (\annotationX,\fourthPlane,\annotationY);

        \def\trackFirstX{0}
        \def\trackFirstY{0}
        \def\trackSecondX{0}
        \def\trackSecondY{0.020373}
        \def\trackPhantomInX{0}
        \def\trackPhantomInY{0.061489}
        \def\trackControlsInX{-0.1}
        \def\trackControlsInY{0.061489}
        \def\trackControlsOutX{-0.1}
        \def\trackControlsOutY{0.1}
        \def\trackPhantomOutX{-0.1}
        \def\trackPhantomOutY{-0.2}
        \def\trackThirdX{-0.5}
        \def\trackThirdY{-0.8}
        \def\trackFourthX{-0.6}
        \def\trackFourthY{-1}
        \def\hitSize{4pt}

        \begin{scope}[canvas is xz plane at y=\firstPlane]
            \fill[TableauPurple] (\trackFirstX,\trackFirstY) circle (\hitSize);
            \draw (\trackFirstX,\trackFirstY) circle (\hitSize);
        \end{scope}
        \begin{scope}[canvas is xz plane at y=\secondPlane]
            \fill[TableauPurple] (\trackSecondX,\trackSecondY) circle (\hitSize);
            \draw (\trackSecondX,\trackSecondY) circle (\hitSize);
        \end{scope}
        \begin{scope}[canvas is xz plane at y=\thirdPlane]
            \fill[TableauPurple] (\trackThirdX,\trackThirdY) circle (\hitSize);
            \draw (\trackThirdX,\trackThirdY) circle (\hitSize);
        \end{scope}
        \begin{scope}[canvas is xz plane at y=\fourthPlane]
            \fill[TableauPurple] (\trackFourthX,\trackFourthY) circle (\hitSize);
            \draw (\trackFourthX,\trackFourthY) circle (\hitSize);
        \end{scope}
        \begin{scope}[canvas is xz plane at y=\phantomStart]
            \node[mark size=2pt,color=TableauPurple] at
            (\trackPhantomInX,\trackPhantomInY) {\pgfuseplotmark{diamond*}};
        \end{scope}
        \begin{scope}[canvas is xz plane at y=\phantomEnd]
            \node[mark size=2pt,color=TableauPurple] at
            (\trackPhantomOutX,\trackPhantomOutY) {\pgfuseplotmark{diamond*}};
        \end{scope}
        \draw[TableauPurple, dotted, thick]
        (\trackFirstX,\firstPlane,\trackFirstY) --
        (\trackSecondX,\secondPlane,\trackSecondY) --
        (\trackPhantomInX,\phantomStart,\trackPhantomInY);
        \draw[TableauPurple, dashed, thick]
        (\trackPhantomInX,\phantomStart,\trackPhantomInY) ..
        controls (\trackControlsInX, \phantomStart+1, \trackControlsInY)
        and (\trackControlsOutX, \phantomEnd-1, \trackControlsOutY) ..
        (\trackPhantomOutX,\phantomEnd,\trackPhantomOutY);
        \draw[TableauPurple, dotted, thick]
        (\trackPhantomOutX,\phantomEnd,\trackPhantomOutY) --
        (\trackThirdX,\thirdPlane,\trackThirdY) --
        (\trackFourthX,\fourthPlane,\trackFourthY);

        \draw[yshift=\hitSize,latex-] (\trackFirstX,\firstPlane,\trackFirstY) --
        (\trackFirstX,\firstPlane,\trackFirstY+0.25);
        \draw[yshift=-\hitSize,latex-] (\trackFirstX,\firstPlane,\trackFirstY) --
        (\trackFirstX,\firstPlane,\trackFirstY-0.25) node[left] {\positionResolution};

        \draw[yshift=\hitSize,latex-] (\trackSecondX,\secondPlane,\trackSecondY) --
        (\trackSecondX,\secondPlane,\trackSecondY+0.25);
        \draw[yshift=-\hitSize,latex-] (\trackSecondX,\secondPlane,\trackSecondY) --
        (\trackSecondX,\secondPlane,\trackSecondY-0.25) node[left] {\positionResolution};

        \draw[yshift=\hitSize,latex-] (\trackThirdX,\thirdPlane,\trackThirdY) --
        (\trackThirdX,\thirdPlane,\trackThirdY+0.25);
        \draw[yshift=-\hitSize,latex-] (\trackThirdX,\thirdPlane,\trackThirdY) --
        (\trackThirdX,\thirdPlane,\trackThirdY-0.25) node[left] {\positionResolution};

        \draw[yshift=\hitSize,latex-] (\trackFourthX,\fourthPlane,\trackFourthY) --
        (\trackFourthX,\fourthPlane,\trackFourthY+0.25);
        \draw[yshift=-\hitSize,latex-] (\trackFourthX,\fourthPlane,\trackFourthY) --
        (\trackFourthX,\fourthPlane,\trackFourthY-0.25) node[left] {\positionResolution};

    \end{tikzpicture}
    }

    \caption{\label{fig:Simulation} Layout of the Monte Carlo based analysis: a subdivided, homogeneous water box with thickness \thickness{} represents the phantom.
        Two pairs of silicon detector slabs with position resolution \positionResolution{} and material budget \materialBudget{} are placed symmetrically upstream and downstream of the phantom.
        The distance \distance{} keeps detectors of a pair apart and a clearance \clearance{} is the gap size between inner detectors and phantom.
        Protons or helium ions from a beam with an initial energy \energy{} are recorded at the detectors and throughout the phantom.}
\end{figure}
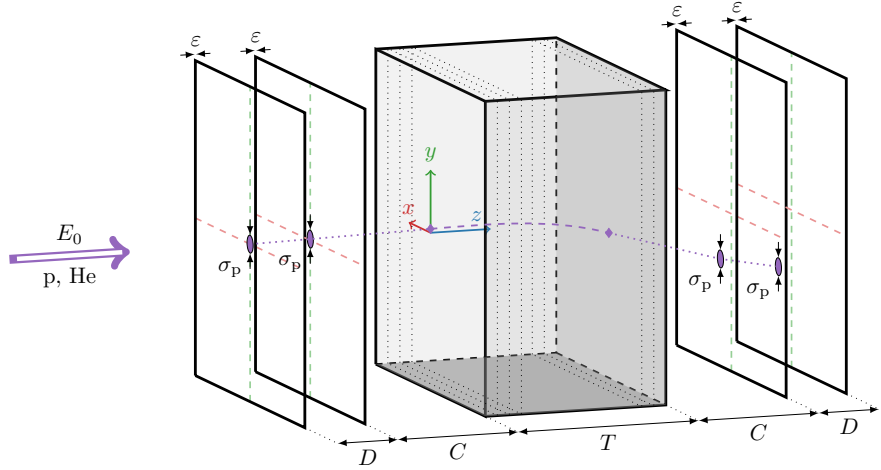

A parameterised Monte Carlo simulation was built with the Geant4 toolkit, version \texttt{10.5.p01}.
The reference physics list \textit{QGSP\textunderscore BIC} was used in most simulations since it provides standard electromagnetic physics processes such as multiple scattering, ion energy loss, Bremsstrahlung and pair production, and elastic and inelastic interactions of hadrons.
One single simulation instead used the \textit{G4EmStandardPhysics\textunderscore option3} physics list -- which did not take hadron interactions into account -- to illustrate the effects of filtering events with nuclear interactions (section \ref{sec:DataFiltering}).
\num{5e5} primary particles were used in each of the simulations. 
A simplified ion imaging layout for single particle tracking, as introduced in \textcite{Schulte2004}, was modeled in the simulations (figure \ref{fig:Simulation}).
The layout consists of four thin detector slabs arranged as one pair upstream of a large phantom and one pair downstream of it.
Each of the volumes constituting the simulation had a fixed \SI{500x500}{\lengthUnit} extent perpendicular to the beam direction.
Detector and phantom thicknesses were configured using parameters which are explained in more detail in section \ref{sec:ParameterValues}.

Hit positions on the detectors were used for the analysis to find the boundary conditions for a path model.
During the simulations, positions and energies were measured with no uncertainty (ideal accuracy) in order to limit the total number of simulations that needed to be carried out.
A finite position resolution of the detectors was added after the simulations, in the subsequent analysis.

The phantom box was subdivided into \SI{1}{\lengthUnit} thin slabs and declared a sensitive detector to sample positions and energies of each particle as it moved through the phantom.
Position samples within the phantom served as ground truth for the evaluation of the \gls{mlp} model.
Energy samples were converted to the kinematic terms \(pv(z)\), needed for the \gls{mlp} model, according to

\begin{equation}
    pv(z) =
    E_\text{kin}(z) \frac{E_\text{kin}(z) + 2 mc^2}{E_\text{kin}(z) + mc^2},
\end{equation}

where \(p\), \(v\) and \(E_\text{kin}\) are the mean momentum, mean velocity and mean kinetic energy of the particles, each at a depth \(z\) within the phantom.
The term \(mc^2\) is the rest energy of the projectile.
Following common practice in ion imaging \parencite{Williams2004,Li2006,Schulte2008}, a curve fit to a fifth order polynomial was used to simplify the integrals underlying the \gls{mlp} model, according to

\begin{equation}
    \frac{1}{p^2v^2(z)} = \sum_{i=0}^5 a_i z^i,
\end{equation}

with coefficients \(a_i\).
Fits were carried out for each individual simulation since the polynomial depends on depth and energy.

\subsection{Boundary conditions}
\label{sec:BoundaryConditions}

Path models are a class of functions that attempt to recreate the original path of a particle undergoing multiple Coulomb scattering in a medium.
These functions take boundary conditions at the medium surface -- the position and direction at the entrance and exit -- as input to model the position at any depth within it.
The simplest model, which disregards the direction at the boundaries, is a linear interpolation from entrance to exit position.
More sophisticated models, such as two straight lines with a single kink \parencite{Jansen2018}, a cubic spline \parencite{CollinsFekete2015} or the most likely path \parencite{Williams2004} are preferential to the linear model, since they produce a more accurate estimate.

\begin{figure}
    \centering
    \includegraphics{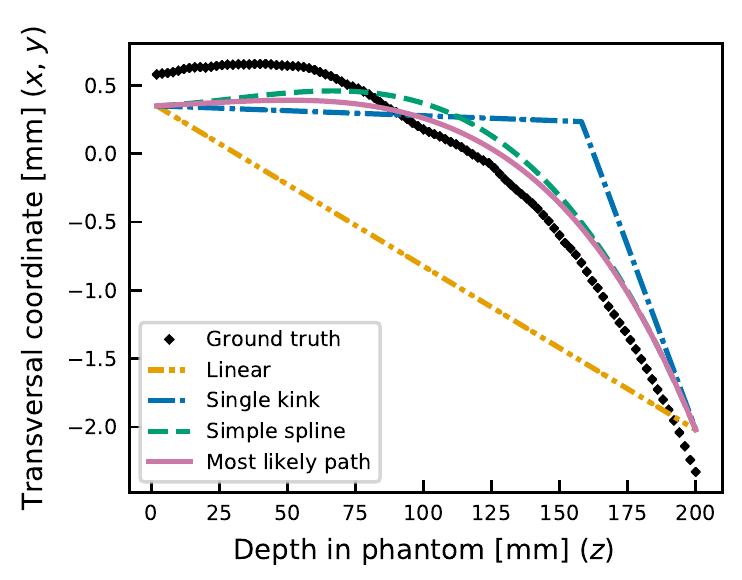}
    \caption{\label{fig:Track} A simulated track of a single proton with several path model estimates. 
        Non-zero differences remain at the entrance and exit due to external detector uncertainties.}
\end{figure}

Three steps were carried out to obtain the boundary conditions for each particle track.
First, \(3\sigma\) cuts on energy loss and scattering angle over the phantom were applied to reduce the dataset to those events that only underwent multiple Coulomb scattering.
By filtering out nuclear interactions and large angle scatter events, the uncertainty within the phantom can be significantly reduced \parencite{Schulte2008}.
Second, a normal distribution with zero mean and the position resolution \positionResolution{} as standard deviation was used to draw random error terms that were added to the \(x\)- and \(y\)-components of each hit location to emulate measurement uncertainty.
Third, the detector hits from each simulation were converted to the boundary conditions for a path model.
A direction vector was calculated as the difference of the two hit positions for each of the entry and exit detector pairs.
Then, the boundary positions at the phantom surface were obtained by propagating hits to the surface in a straight line along the direction vector.
These boundary conditions did not generally reproduce the real positions at the surface due to the added error terms and scattering in the detectors and the surrounding air, and a deviation remained (figure \ref{fig:Track}).

\begin{figure}
    \centering
    \includegraphics{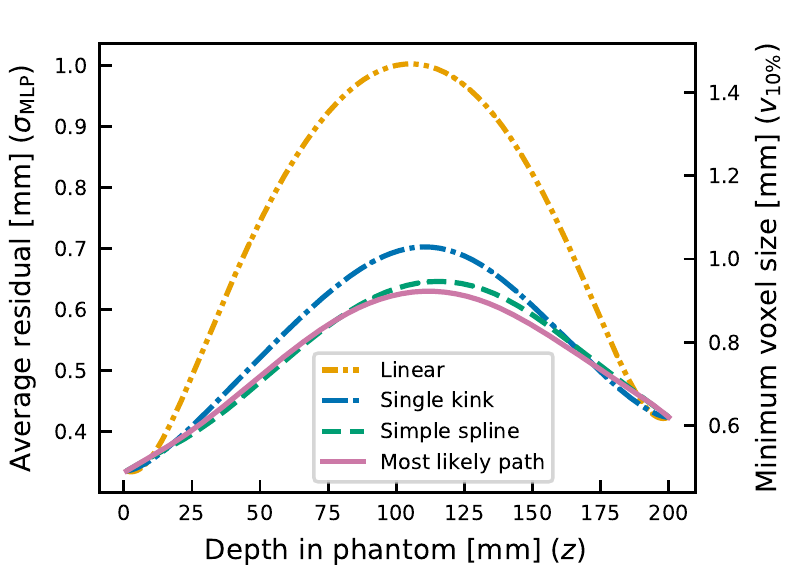}
    \caption{\label{fig:ModelRms} The root-mean-square deviation of transversal coordinates \uncertainty{} as function of depth.
        An additional axis on the right shows the estimated minimum pixel size \(v_{10\%}\) corresponding to \uncertainty{}.}
\end{figure}

\subsection{Uncertainty envelope}
\label{sec:UncertaintyEnvelope}

The \gls{mlp} model was sampled at those \(z\)-positions in the phantom that were previously recorded as ground truth during simulations, and the difference between model and ground truth was calculated for each particle at each depth.
Because the distributions of the \(x\)- and \(y\)-components are uncorrelated, both sets were combined to obtain a single distribution with twice the size.
The \gls{rmsd} at those positions was used to summarise over all events of a simulation and to describe the \gls{mlp} uncertainty \(\sigma_\mathrm{MLP}(z)\) (figure \ref{fig:ModelRms}).

Shape and height of each uncertainty envelope depended on the parameters of the simulation and the path model used.
Within the phantom, the uncertainty usually increased towards the middle, due to a rising distance from the known positions on the detectors.
A maximum was often observed in the second half of the phantom because of the energy loss over the phantom.
Another consequence of this energy loss is that the uncertainty was generally higher at the exit than at the entrance.

In this work, the figure of merit is based on the spatial frequency \(f_{10\%}\), at which the modulation transfer function decays below \SI{10}{\percent}, according to the findings in \textcite{Krah2018}.
\(f_{10\%}\) was calculated as a function of the uncertainty and, when interpreted as the Nyquist frequency of a system, allowed to obtain a lower limit on the useful image pixel spacing \(v_{10\%}\)

\begin{equation}
    v_{10\%} =
    \frac{0.5}{\min\left(f_{10\%}(z)\right)} =
    \frac{\pi}{\sqrt{2\ln 10}} \times \max \left(\sigma_\text{MLP}(z)\right).
\end{equation}

For any given system, the maximum value of the uncertainty envelope corresponded to a minimum in spatial frequency.
The pixel spacing at this frequency was used to find a conservative minimum pixel spacing as an intuitive figure of merit.
It should be noted that uncertainty due to image reconstruction methods was not taken into account, and realistic image resolutions from measurements can likely be lower than those that are reported in this work.

\subsection{Parameter values}
\label{sec:ParameterValues}

To investigate the influence of system parameters on the accuracy of path models, simulations were carried out with different parameter sets.
While the extent of the volumes perpendicular to the beam and their materials were constant -- sensor slabs, the phantom and the surrounding volume were made of silicon, water and air, respectively -- several other parameters were variable.
Parameters and their value intervals are summarised in table \ref{tab:parameterValues} and are explained in the following paragraphs.

The system geometry was defined by the detector distance \distance{} and the clearance \clearance{} between inner detectors and the phantom.
Detector distance was expected to improve path accuracy for larger distances because of a reduced angular uncertainty due to position resolution \parencite{Poludniowski2015}.
Previous studies in proton imaging suggested to keep the distance above at least \SI{80}{\lengthUnit} upstream and \SI{60}{\lengthUnit} downstream \parencite{Penfold2011} or both above \SI{50}{\lengthUnit} \parencite{Bopp2014}, and similar findings were expected to be observed with helium ions instead of protons.

Clearance between phantom and the inner trackers was also expected to increase the uncertainty due to scattering in the detectors and position resolution.
Contrary to distance, clearance should be kept as low as possible because a larger gap size was expected to reduce the accuracy \parencite{Schneider2012,Sadrozinski2013}.
However, a minimum clearance between \SIrange{100}{300}{\lengthUnit} may be necessary for safety and technical reasons \parencite{Schulte2004}.

The two parameters that described the tracking detectors were the position resolution \positionResolution{} and its thickness, given as material budget \materialBudget{}.
Position resolution was the spread of random error terms added to the tracker hit positions, which reduced the accuracy of the boundary conditions on the phantom surface.
Thus, a tracker with a finer position resolution was expected to produce more accurate path estimates.
Proton CT scanners usually utilise sensors with a pitch in the range of \SIrange{90}{500}{\positionResolutionUnit} \parencite{Sadrozinski2013,Taylor2015,Uzunyan2016}, or some as small as \SI{28}{\positionResolutionUnit} \parencite{Mattiazzo2018}.
The detector's thickness was described by its total material budget, i.e. the sum of material thicknesses divided by their corresponding radiation length for each material traversed by the particles.
Detectors with a larger material budget were expected to perform worse than thinner ones, due to the increased amount of scattering in the detector itself.
Besides the sensor material itself, electronics, data lines, services, powering, cooling and mounting contribute to (and often dominate) the total material budget of a detector.
For reasons of simplicity, only a single silicon slab was used per detector plane in the simulations underlying this work.
The considered values represent typical values in existing and future high-energy physics detectors such as the CMS Outer Tracker \parencite{CMSOTtdr}, the ALICE Inner Tracking System \parencite{ALICEITStdr} and the CLIC vertex detector \parencite{CLICdet}. 

Finally, the beam parameters were the particle species \(p\) and its primary energy \energy{}, both of which influenced the amount of scattering \parencite{Moliere1948}.
Increasing the beam energy was expected to improve path reconstruction.
Energies in the range of \SIrange{200}{350}{\energyUnit} are usually studied for a head-sized phantom \parencite{Schulte2004,CollinsFekete2017,Johnson2017}.
The standard deviation of the \gls{mlp} estimate due to ion multiple scattering alone scales by a factor of \((z/A)^2\) \parencite{CollinsFekete2017}, where \(z\) is the particle's charge and \(A\) its mass number.
It was thus expected that the (intrinsic) uncertainty for helium is approximately one quarter of the uncertainty for protons.
Indeed, \textcite{CollinsFekete2017} demonstrated with a Monte Carlo simulation that the intrinsic maximum \gls{rmsd} of helium is about one third of that for protons given a fixed initial energy of \SI{350}{\energyUnit} for both particles.
Carbon ions were not considered for this work, since radiographic images based on carbon require a dose exceeding clinical reference values, a disadvantage that does not affect protons or helium ions \parencite{Gehrke2018}.

\begin{table}
    \centering
    \small
    \caption{\label{tab:parameterValues} Summary of the parameters and their range of values used for the Monte Carlo-based analysis.}
    \begin{tabular}{l l l}
        \hline
        Variable                & Name                & Parameter range                                     \\
        \hline
        \distance{}             & Detector distance   & \SIrange{25}{300}{\lengthUnit}                      \\
        \clearance{}            & Phantom clearance   & \SIrange{100}{300}{\lengthUnit}                     \\
        \thickness{}            & Phantom thickness   & \SIrange{100}{300}{\lengthUnit}                     \\
        \positionResolution{}   & Position resolution & \SIrange{1}{250}{\positionResolutionUnit}           \\
        \materialBudget{}       & Material budget     & \SIrange{0.001}{2}{\materialBudgetUnit} (\(x/X_0\)) \\
        \energy{}               & Beam energy         & \SIrange{250}{500}{\energyUnit}                     \\
        \(p\)                   & Particle species    & Proton, helium                                      \\
        \hline
    \end{tabular}
\end{table}

\section{Results}
\label{sec:Results}

The beneficial effects of \(3\sigma\)-cuts are demonstrated at different position resolutions, for protons (section \ref{sec:DataFiltering}).
Detector distance was found to weakly interact with other parameters and is discussed separately (in section \ref{sec:DetectorDistance}).
Similarly, beam energy is discussed separately, because the benefits of increased energy were found to be limited (section \ref{sec:BeamEnergy}).
Finally, the pixel spacing is presented as a function of detector material budget and position resolution, for combinations of other parameter values (section \ref{sec:DetectorProperties}).
This overview can be used to delimit intervals of position resolution and material budget that are useful for ion imaging in general or for high-resolution imaging applications.

\subsection{Data filtering}
\label{sec:DataFiltering}

\begin{figure*}
    \centering
    \includegraphics[width=\linewidth]{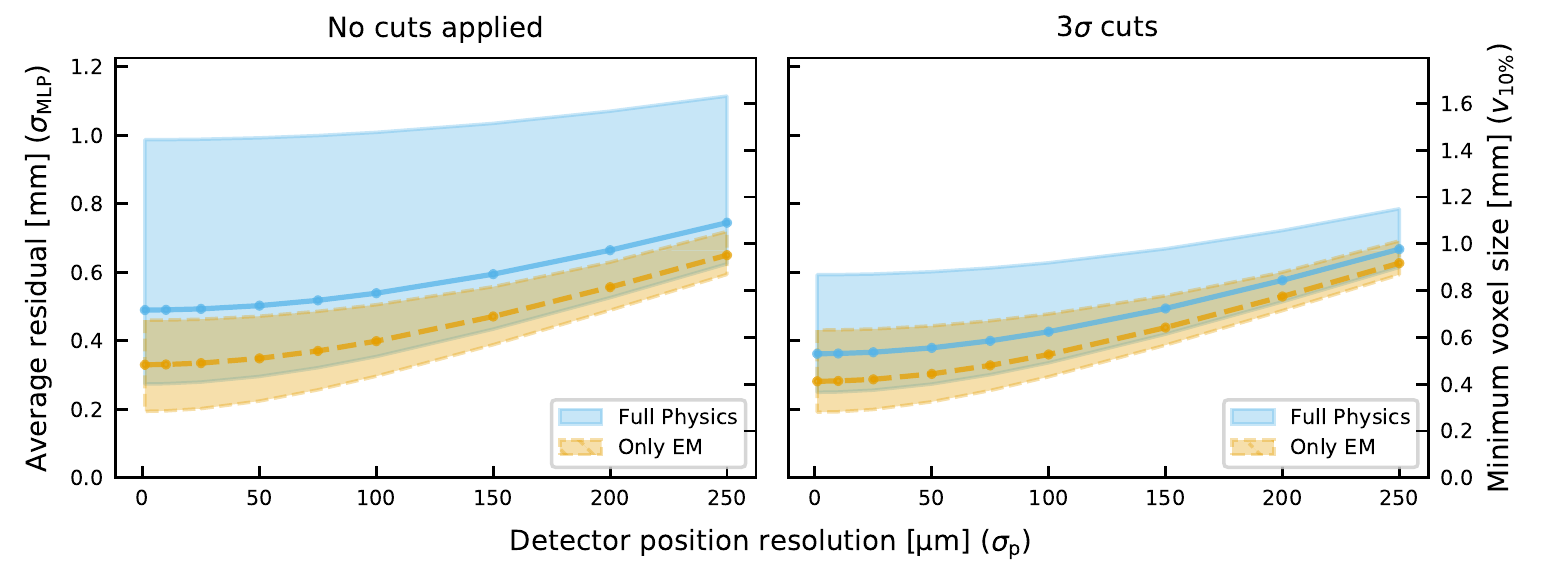}
    \caption{\label{fig:filtering} Effects of filtering at different position resolutions for two physics lists, using protons and a \SI{200}{\lengthUnit} phantom size.
        Data with no cuts applied (left) were compared to \(3 \sigma\) cuts on scatter angle and energy loss (right).
        Filled areas represent the minimum (i.e. entrance) and maximum \gls{rmsd} along the phantom depth and lines represent the \gls{rmsd} at the phantom exit.}
\end{figure*}



Normally, events with scatter angles and energy losses beyond \(3\sigma\) of their respective distributions were removed from the analysis to improve the path model \gls{rmsd} (section \ref{sec:BoundaryConditions}).
This improvement is illustrated using results from two simulations with protons traversing a \SI{200}{\lengthUnit} phantom. 
One simulation used the reference physics list \textit{QGSP\textunderscore BIC} and was labelled \textit{Full Physics}, while the other simulation used the \textit{G4EmStandardPhysics\textunderscore option3} physics list and was labelled \textit{Only EM}.
Both simulations were analysed with and without cuts applied, at different position resolutions (figure \ref{fig:filtering}).

\(3\sigma\) cuts provided no benefit for the \textit{Only EM}-simulation, because the resulting distribution of scattering angles was based only on the central Gaussian approximation to the Moli\'ere model \parencite{Lynch1991, Geant4PhysicsReference}.
Though the cut removed some events, it had no impact on the shape of the scattering distribution.
However, additional large angle scatter events occurred in the \textit{Full Physics} simulation, which were partially filtered out by the cuts.
In the presented case (figure \ref{fig:filtering}) the cuts reduced the maximum \gls{rmsd} (top of the blue band) as well as the \gls{rmsd} at the exit (blue line).
Because nuclear events mainly occurred in the phantom, no such improvement was observed for the entrance uncertainty (bottom of the blue band).

Results also showed that the potential improvements in path uncertainty decreased with a finer position resolution.
This is because other parameters and the intrinsic scattering became dominant.

\subsection{Detector distance}
\label{sec:DetectorDistance}

\begin{figure*}
    \centering
    \includegraphics[width=\linewidth]{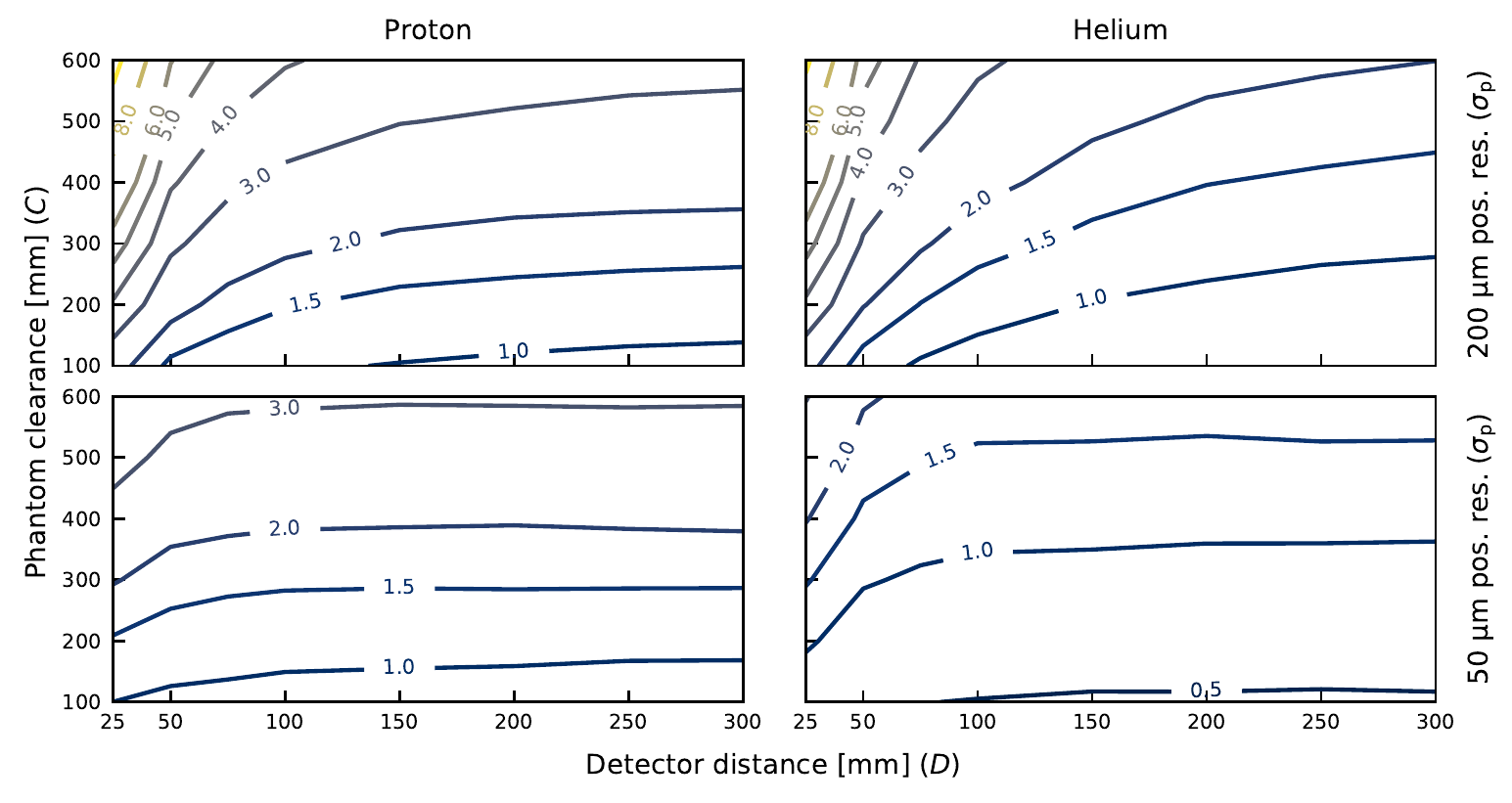}
    \caption{\label{fig:distanceClearance} Pixel spacing as function of detector distance and phantom clearance, for protons (left) and helium ions (right).
    Two position resolutions were considered: \SI{50}{\positionResolutionUnit} (bottom) and \SI{200}{\positionResolutionUnit} (top).
        Labelled contour lines represent a constant level of pixel spacing in \si{\lengthUnit}.
        Phantom thickness, material budget and beam energy were kept at \SI{200}{\lengthUnit}, \SI{0.4}{\materialBudgetUnit} and \SI{250}{\energyUnit}, respectively.}
\end{figure*}

The distance between detectors was found to only weakly interact with other parameters.
This was illustrated for simulations with a fixed phantom size, material budget and beam energy of \SI{200}{\lengthUnit}, \SI{0.4}{\materialBudgetUnit} and \SI{250}{\energyUnit}, respectively.
In the presented case, phantom clearance was varied between \SIrange{100}{600}{\lengthUnit}, detector distance from \SIrange{25}{300}{\lengthUnit} and two position resolutions -- \SI{50}{\positionResolutionUnit} and \SI{200}{\positionResolutionUnit} -- were considered for protons and helium ions.

Results from these simulations were summarised as labelled contour lines that mark where a fixed pixel spacing in \si{\lengthUnit} lies within the parameter space (figure \ref{fig:distanceClearance}).
A common trend was observed in all four combinations of particle species and position resolution.
At small distances, the contours were more dense and driven by both distance and clearance.
Upwards of detector distances between \SIrange{75}{150}{\lengthUnit} the lines became more sparse and remained nearly parallel to the x-axis.
Uncertainty was mainly driven by phantom clearance in this region, and an increased distance did not effectively improve image resolution.

Considering these findings, the remaining simulations were carried out with a constant detector distance of \SI{100}{\lengthUnit} as a compromise between compactness and accuracy.

\subsection{Beam energy}
\label{sec:BeamEnergy}

\begin{figure*}
    \centering
    \includegraphics[width=\linewidth]{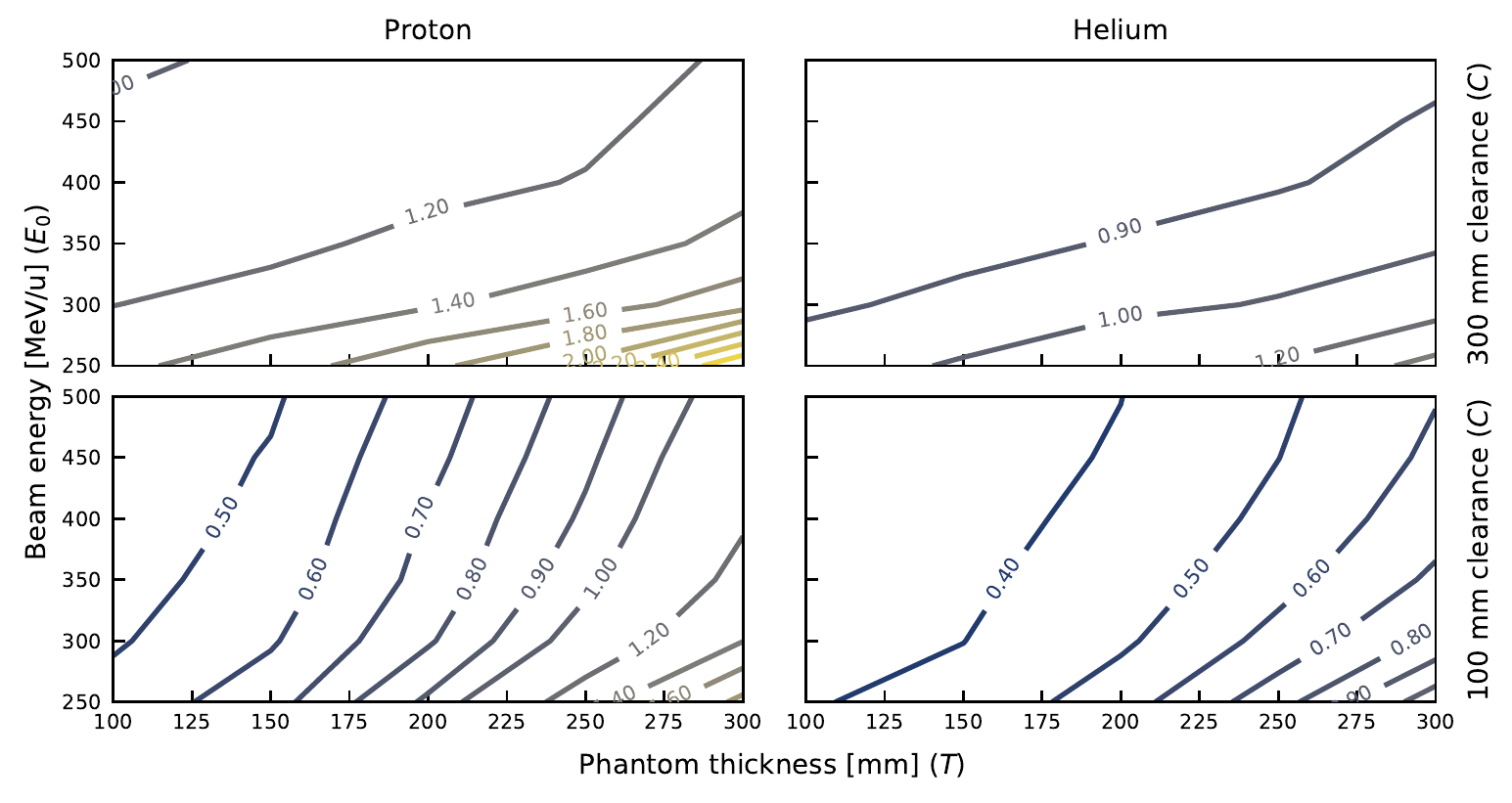}
    \caption{\label{fig:energyThickness}
        Pixel spacing as a function of beam energy and phantom thickness, for protons (left) and helium ions (right).
        Two clearances were considered: \SI{100}{\lengthUnit} (bottom) and \SI{300}{\lengthUnit} (top).
        Labelled contour lines represent a constant level of pixel spacing in \si{\lengthUnit}.
        Energy, position resolution and material budget were kept at \SI{250}{\energyUnit}, \SI{100}{\positionResolutionUnit} and \SI{0.4}{\materialBudgetUnit}, respectively.
        Note that the step size between consecutive levels changes from \SIrange{0.1}{0.2}{\lengthUnit} at \SI{1}{\lengthUnit} in this graph.}
\end{figure*}

The influence of beam energy on path uncertainty was studied using energies between \SIrange{250}{500}{\energyUnit}, for phantom thicknesses from \SIrange{100}{300}{\lengthUnit}.
Position resolution and material budget were kept at \SI{100}{\positionResolutionUnit} and \SI{0.4}{\materialBudgetUnit} respectively, and two different clearances of \SI{100}{\lengthUnit} and \SI{300}{\lengthUnit} were used.
Contours revealed how an increase in energy reduced the minimal pixel size, especially at a large clearance (figure \ref{fig:energyThickness}).

Additional energy was found to improve pixel size for protons and helium ions in the same way, however contours were less dense when helium was used.
With increasing energy, the potential for additional improvement was reduced for both particles.
Using additional energy was also less effective at a smaller phantom clearance.
In these cases the contours were lower, less dense and steeper in the thickness-energy plane.

For protons in particular the contour density was elevated at lower energies and large phantom thicknesses, as can be seen in the upper left panel of figure \ref{fig:energyThickness}.
In this region the pixel spacing for a \SI{200}{\lengthUnit} thick phantom could be improved from \SIrange{1.75}{1.4}{\lengthUnit} simply by increasing the proton energy from \SIrange{250}{300}{\energyUnit} (see upper left panel in figure \ref{fig:energyThickness}).

\subsection{Position resolution and material budget}
\label{sec:DetectorProperties}

\begin{figure*}
    \centering
    \includegraphics[width=\linewidth]{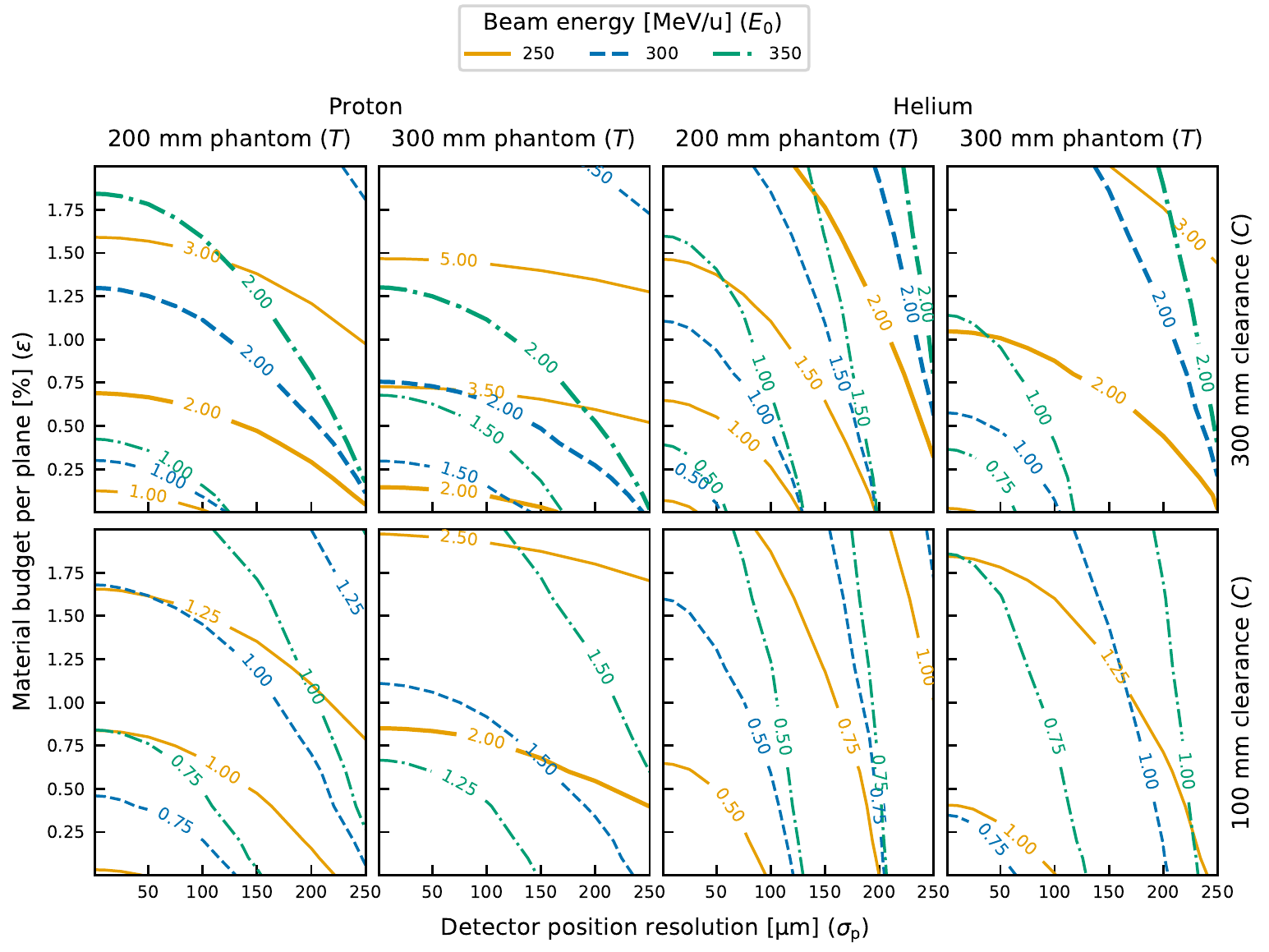}
    \caption{\label{fig:parameterSpace} Pixel spacing in \si{\lengthUnit} as function of position resolution, material budget and energy.
        Rows are mapped to phantom clearance (\SIrange{100}{300}{\lengthUnit}) and columns to both phantom thickness (\SIrange{200}{300}{\lengthUnit}) and beam particle species (proton, helium).
        To be suitable for a target pixel spacing, a system's material budget and position resolution should be below and to the left of the corresponding contour line. }
\end{figure*}

The influence of the detector properties on the achievable pixel spacing was investigated for many combinations of the other parameters.
Results were grouped in terms of beam particle, phantom thickness and clearance to create an overview over the parameter space (figure \ref{fig:parameterSpace}).
Phantom thicknesses of \SI{200}{\lengthUnit} and \SI{300}{\lengthUnit} were considered for the overview, as well as clearances of \SI{100}{\lengthUnit} and \SI{300}{\lengthUnit}.

The contours gradually flattened in the position resolution-material budget plane, especially below a position resolution of \SIrange{50}{100}{\positionResolutionUnit}.
This means that only small gains were possible for high resolution sensors.
Improvements in terms of position resolution played a slightly larger role at an increased beam energy (different colour contours in figure \ref{fig:parameterSpace}), where the slope in the position resolution-material budget plane was steeper and the density of contour lines was increased slightly.
At higher energies the contours were moved towards a slightly worse position resolution, especially for the thicker phantom (second and fourth column), indicating that energy could be used to compensate for a lack of position resolution.
Position resolution was more relevant at a larger clearance between the inner detectors and the phantom.
For a clearance of \SI{300}{\lengthUnit} (top row in \ref{fig:parameterSpace}) the contour density was increased in the direction of position resolution, compared to a clearance of \SI{100}{\lengthUnit} (bottom row).
Additionally, the contours were shifted towards lower position resolution values.
Overall the largest influence of position resolution on the image resolution was observed for a helium beam and a \SI{200}{\lengthUnit} phantom at a \SI{300}{\lengthUnit} clearance.

In contrast to position resolution, a reduction in detector material budget improved the path uncertainty even for thin sensors.
The influence of material budget on path uncertainty strongly depended on beam energy and particle species.
When increasing the energy, contour levels moved towards larger material budget values and their density decreased, because the amount of scattering was reduced.
It is evident in all graphs of figure \ref{fig:parameterSpace} that the slope of contours changed to be more parallel to the material budget axis and the levels became more sparse.
In the examined parameter space, the contour density increased more quickly in terms of material budget than in terms of position resolution.
Material budget was also found to strongly depend on phantom clearance.
For protons (left two columns in figure \ref{fig:parameterSpace}) the density was roughly six times larger at \SI{300}{\lengthUnit} than at \SI{100}{\lengthUnit}.

Pixel spacing levels were found to be consistently smaller for helium ions compared to protons, given otherwise identical parameters.
The minimum pixel spacing was lower due to the reduced amount of scattering that the heavier helium ion was subjected to.
A change from protons to helium ions also impacted the influence of other system parameters.
Section \ref{sec:BeamEnergy} describes how an increase in beam energy improved the path uncertainty of helium in the same way that it did for protons, however at a lower rate.
Similar observations are displayed in figure \ref{fig:parameterSpace}.
The contours of different energies were closer together for helium than for protons, and the contours were also steeper, because material budget was a less deciding factor.
An interesting feature in the top row of figure \ref{fig:parameterSpace} is that helium contours often intersected the y-axis at a similar value of material budget, where a contour level twice as large was present for protons.
Thus, for high resolution sensors, the disadvantage of using a thick sensor over a thin one is reduced by half if a helium beam is available.

\section{Discussion}
\label{sec:Discussion}

The aim of this work was to create an overview of the minimum achievable pixel spacing of single particle tracking ion imaging systems, by using Monte Carlo simulations.
This overview was intended to guide future hardware developments, by taking detector parameters, geometry and beam properties into account.
Detector hits were distorted with randomly generated error terms to emulate position resolution and the most likely path model was compared to the actual path for each simulated event.
One depth-dependent uncertainty envelope per simulation was obtained as the root-mean-square deviation between model and ground truth.
The maximum along this curve was converted to the spatial frequency at which the corresponding modulation transfer function decays below \num{0.1} and then used as Nyquist frequency to evaluate a conservative pixel spacing.

For detector properties typically used in ion imaging -- i.e. a material budget below \SI{0.75}{\materialBudgetUnit} and position resolutions of less than \SI{150}{\positionResolutionUnit} -- lower limits on pixel spacing were found between \SIrange{1}{2}{\lengthUnit} for protons and \SIrange{0.5}{1.5}{\lengthUnit} for helium.
Similar findings were obtained in other Monte-Carlo based studies, which reported spatial frequencies between \SIrange{0.25}{0.8}{\imageResolutionUnit} \parencite{Li2006, Plautz2014, Hansen2014, Plautz2016, Krah2018}, corresponding to pixel spacing values of \SIrange{2}{0.625}{\lengthUnit}.

\subsection{System geometry}

Results indicated that a constant distance between the detectors of a pair could be used in the analysis, because it barely influenced the effects of other parameters.
Path accuracy degraded more quickly below a distance of \SI{50}{\lengthUnit} and remained almost constant above \SI{150}{\lengthUnit}.
Ideally this distance should be large, however a compact system with \SI{100}{\lengthUnit} was found to be adequate already.
This is in line with similar observations from previous studies using Monte Carlo simulations \parencite{Penfold2011} or an analytical method \parencite{Bopp2014}.

While detector distance should be kept reasonably large, phantom clearance should be as small as reasonably achievable.
In the conducted simulations it was found that the clearance between phantom and inner detectors amplified uncertainty due to scattering and position resolution and should, hypothetically, be close to zero.

\subsection{Treatment planning requirements}

Clinical treatment plans traditionally based on X-ray CT images work with a slice thickness of \SI{2}{\lengthUnit} or \SI{3}{\lengthUnit}, depending on the tumour location.
A better image resolution of \SI{1}{\lengthUnit} is required to properly predict the dose to small organs at risk, like for example the optical nerves, chiasm or the cochlea, especially for the treatment with charged particles such as protons and light ions.
Sparing these organs is the rationale for using particle therapy, especially with respect to quality of life for patients.
A hypothetical single particle tracking system must be at least accurate enough to fulfil basic clinical requirements.

Regions of parameters that fulfilled such requirements can be identified in figure~\ref{fig:parameterSpace}.
A pixel size of \SI{2}{\lengthUnit} could be achieved for a given beam energy when the detector material budget was below, and the position resolution to the left of the corresponding contour line.
At a clearance of \SI{100}{\lengthUnit} this was generally fulfilled, with a single limitation in that the material budget should be below \(\approx\)\SIrange{0.5}{0.75}{\materialBudgetUnit} for protons irradiating a \SI{300}{\lengthUnit} phantom.
A slight increase in beam energy was enough to remove this limitation, since the \SI{2}{\lengthUnit} contour level for \SI{300}{\energyUnit} was already beyond the scales.

Material budget and position resolution values below the \SI{2}{\lengthUnit} contour were more restricted at a large clearance of \SI{300}{\lengthUnit}.
A material budget between \SI{0.25}{\materialBudgetUnit} for a position resolution of \SI{200}{\positionResolutionUnit} and \SI{0.75}{\materialBudgetUnit} for a position resolution of \SI{10}{\positionResolutionUnit} should be used for protons irradiating a \SI{200}{\lengthUnit} phantom representing the dimension of an adult head.
A similar restriction applies to protons irradiating a \SI{300}{\lengthUnit} phantom, which is at least needed for irradiations in the pelvic region, provided that the beam energy is increased from \SIrange{250}{300}{\energyUnit}.
Detector properties were generally found less restricted with a helium beam.
For a \SI{300}{\lengthUnit} phantom at a large clearance of \SI{300}{\lengthUnit}, the material budget should be kept below \SI{0.5}{\materialBudgetUnit} for a position resolution of \SI{200}{\positionResolutionUnit} and \SI{1}{\materialBudgetUnit} for a position resolution of \SI{10}{\positionResolutionUnit}.

\subsection{Increasing the image resolution}

While a \SI{2}{\lengthUnit} pixel spacing is suitable for standard clinical cases, other applications could potentially benefit from a more refined grid.
In particle therapy, the presence of implants, surgical clips or markers must be handled carefully.
A finer CT resolution and dose calculation grid of \SI{1}{\lengthUnit} or less improves the dose calculation and reduces the dose prediction uncertainties \parencite{Righetto2020, Jia2015}.
More complicated indications in the thorax, such as in the lung region, also raise the need for image resolution in the sub-millimetre range.
The sponge-like structure of lung tissue cannot be visualised with a resolution below \SI{1}{\lengthUnit} but has an impact on the beam characteristics, and therefore on the dose prediction \parencite{Espana2011,Baumann2019,Hranek2020}.
In addition, respiratory motion poses the need for motion determination and tracking with a high spatial and temporal resolution, using time resolved 4D imaging data.
The low resolution of these imaging sets can lead to imaging artefacts as well a lack of structural information \parencite{Fang2017}.
Daily control imaging gains more and more importance, especially in particle therapy where a high position accuracy is essential.
A resolution of at least \SI{1}{\lengthUnit} is the pre-condition to detect positioning deteriorations during treatment that can cause a relevant shift in the high and low dose regions \parencite{Ricotti2020}. 

With protons and a \SI{200}{\lengthUnit} phantom size, a sub-millimetre spacing is in reach under certain conditions.
For a small clearance and \SI{250}{\energyUnit} the material budget should be kept below \(\approx\)\SIrange{0.25}{0.75}{\materialBudgetUnit}, depending on position resolution, and the position resolution must be better than \SI{200}{\positionResolutionUnit}.
In this scenario the restrictions on material budget and position resolution may be lifted by increasing the energy.

No sub-millimetre pixel spacing was observed for the \SI{300}{\lengthUnit} phantom when using protons.
This limits the use of protons in the thorax region, where a high image resolution is of increased interest for moving targets.
A more promising option for such an application are helium ions, since high-resolution images are more easily obtained with such a beam.
Sub-millimetre pixel sizes can be attained below a position resolution of \SI{100}{\positionResolutionUnit} and a material budget less than \SI{0.5}{\materialBudgetUnit}, even for a \SI{300}{\lengthUnit} clearance.
Again, the restriction on material budget can be lifted by using a higher energy beam.
The smallest pixel spacing values -- which were below half a millimeter -- were observed using a helium beam at low clearance and a small phantom size.
Although the contour lines gradually move towards thinner and more accurate sensors when the clearance is increased, they do not completely disappear from the graph, which means that an adequate grid for helium CT of a head is in the range \SIrange{0.5}{1}{\milli\meter}.
Conversely, the adequate grid size for protons was found to be just above \SI{1}{\milli\meter}.

\section{Conclusion}
\label{sec:Conclusion}

A comprehensive comparison of single particle tracking system parameters and their influence on the achievable image resolution was carried out in this work, based on Monte-Carlo simulations.
Regions in the parameter space that fulfil basic requirements for treatment planning in ion therapy were identified.
Tracker material budget and position resolution that allow imaging with a pixel spacing below \SI{2}{\lengthUnit} were of particular interest for various combinations of particle species, energy and clearance.
The material budget per tracking plane should remain below \SIrange{0.25}{0.75}{\materialBudgetUnit} for position resolutions from \SIrange{200}{10}{\positionResolutionUnit} when a large clearance is used during proton imaging.
In helium imaging the material budget should not exceed \SIrange{0.5}{1}{\materialBudgetUnit} per plane to be suitable for treatment planning.

The possibilities for imaging with a higher resolution were explored in addition to fulfilling the clinical requirements.
A sub-millimetre pixel spacing was observed for proton beams and a head sized phantom only for a low clearance.
Additionally, no sub-millimetre interval has been observed for a phantom size of \SI{300}{\lengthUnit}, limiting the use of protons in the thorax region.
\gls{mlp} uncertainty was consistently lower for helium ions, compared to protons, and the application of helium ions is more promising for high resolution imaging.
A sub-millimetre pixel size could be achieved with a position resolution below \SI{100}{\positionResolutionUnit} and a material budget below \SI{0.75}{\materialBudgetUnit}, even for the larger phantom and a large clearance.
For a head sized phantom, the smallest pixel spacing was observed to be just below \SI{0.5}{\lengthUnit}.
Thus, the superior image resolution when using helium ions could potentially benefit indications in the thorax region. 

Finally, it should be pointed out that pixel spacing values from the analysis at hand merely present a simple approximation of a lower limit.
Patients or phantoms in an ion imaging context do not consist of a homogeneous body of water but rather mixtures of various biological materials.
Additional uncertainty due to material inhomogeneities could shift the contours towards a smaller material budget and position resolution, and should be investigated in a future study.
This also applies to other sources of uncertainty that were not taken into account, such as image reconstruction algorithms.

\section*{Conflicts of interest}

The authors declare no conflict of interest.

\section*{Acknowledgements}
This project received funding from the Austrian Research Promotion Agency (FFG), grant numbers 875854 and 869878.

\section*{References}

\bibliography{Paper.bib}

\end{document}